\documentclass[twocolumn,aps,prc]{revtex4}
\usepackage{float}
\usepackage{graphicx}
\usepackage{dcolumn}
\usepackage{bm}
\usepackage{xcolor}
\usepackage{amsmath} 

\begin{document}



\title{Impact of Initial-State Nuclear and Sub-Nucleon Structures on Ultra-Central Puzzle in Heavy Ion Collisions}

\author{Qi Wang$^1$} 

\author{Long-Gang Pang$^{1}$}
\email[]{lgpang@ccnu.edu.cn}

\author{Xin-Nian Wang$^{1}$}
\email[]{xnwang@ccnu.edu.cn}

\affiliation{$^1$Key Laboratory of Quark and Lepton Physics (MOE) and Institute of Particle Physics, Central China Normal University, Wuhan 430079, China}

\begin{abstract}

Hydrodynamic models fail to describe the near-equal $v_2/v_3$ ratio observed in ultra-central heavy-ion collisions, despite their success in other centrality classes. This discrepancy stems from shear viscosity suppressing higher-order geometric eccentricities, resulting in underestimated $v_3$ when using the conventional QGP viscosity coefficient. We explore two initial-state modifications to resolve this puzzle: (1) enforcing a minimum nucleon separation distance to homogenize distributions, and (2) amplifying sub-nucleon structures to reduce initial eccentricity. Using TRENTo initial conditions and 3+1D viscous hydrodynamic model CLVisc, both approaches significantly lower geometric eccentricity, reduce required viscosity, and narrow the $v_2$–$v_3$ gap in ultra-central collisions. Our results implicate initial-state nuclear and sub-nucleon structures as critical factors in addressing this puzzle. Resolving it would advance nuclear structure studies and improve precision in extracting QGP transport coefficients (e.g., shear viscosity), bridging microscopic nuclear features to macroscopic quark-gluon plasma properties.  

\end{abstract}

\maketitle

\section{Introduction}


The quark-gluon plasma, a state of matter characterized by extremely high temperatures and energies, has been produced in laboratory settings through the use of high-energy heavy-ion collisions at the Relativistic Heavy Ion Collider (RHIC) and the Large Hadron Collider (LHC). This state of matter is analogous to the conditions that prevailed just microseconds after the Big Bang, earning it the metaphorical title of a "little bang" \cite{mmm,yagi2005quark,JohnEllis_2006}.


The collective flow is a phenomenon observed in heavy-ion collision experiments, where particles produced in the collision move preferentially in certain directions, indicating a collective motion of particles in the created medium. This phenomenon provides important information about the properties and behavior of the hot and dense matter \cite{PhysRevLett.105.252302,PhysRevLett.107.032301,ALICE:2014wao}.
It is typically studied through the measurement of the anisotropic parameters of final-state particles in momentum space. The distribution of final-state particles in momentum space can be expressed as a Fourier series expansion in terms of azimuthal angles:
\begin{equation}
    \frac{d N}{d \phi} \propto 1 + 2 \sum_{n=1} v_n \cos[n(\phi - \Psi_n)].
\end{equation}
where $\Psi_n$ corresponds to the symmetry plane angle of order $n$. The dominant flow coefficient in non-central heavy-ion collisions is the second flow harmonic ($v_2$), called elliptic flow. 

Based on the 3+1-dimensional viscous hydrodynamics model CLVisc, it can accurately describe the coefficients of various orders of collective flow across most centrality ranges \cite{Ding_2021,PhysRevC.97.064918,PANG2016272,Pang:2015zrq,PhysRevC.91.044904,PhysRevC.82.014903,PhysRevC.95.014906,Wu:2021fjf}. However, there is a puzzle that the data of the elliptic flow $v_2$ and the triangular flow $v_{3}$ given by the hydrodynamics model are inconsistent with the experimental data in the ultra-central collisions \cite{Denicol:2014ywa}. And the flow coefficients from the two-particle cumulant method $v_2$ and $v_3$ were reported to be almost the same in ultra-central collisions at LHC \cite{CMS:2013bza, PhysRevLett.107.032301, PhysRevC.86.014907} . Numerous efforts have been made to resolve the puzzle from different aspects, such as improved descriptions of initial conditions \cite{PhysRevC.82.041901,PhysRevC.100.024905,PhysRevC.103.024906,WOS:000473829600021,PhysRevC.92.014901,PhysRevC.102.054905,WOS:000574142700002,WOS:001319552800001,PhysRevC.94.024914}, effects of the transport coefficients \cite{PhysRevC.110.L031901,ROSE2014926,LUZUM2013377c,PhysRevC.107.044907}, hydrodynamic fluctuations \cite{KUROKI2023137958,WOS:000377771500002,PhysRevLett.117.182301}, and the equations of state \cite{PhysRevC.98.034909}. However, none of these approaches have yet succeeded in fully explaining the experimental observations of $v_2\{2\}$ and $v_3\{2\}$ within the framework of hydrodynamic-based dynamical models.

In recent years, the study of nuclear structure has garnered significant attention in the context of heavy-ion collisions. Notable advancements have been achieved in various areas, including deformed nuclei \cite{PhysRevLett.124.202301,PhysRevLett.127.242301,Dimri:2023wup,PhysRevLett.133.192301,PhysRevC.110.034907,Xu:2024bdh} $\alpha$ clusters \cite{Ding_2023,YuanyuanWang:2024sgp,BIJKER2017154}, neutron halos \cite{Zhou:2009sp,WOS:000290725300030}, and neutron skins \cite{PhysRevLett.125.222301,PhysRevLett.126.172502,Giacalone:2023cet,Liu:2023pav}. To address the challenges in this field, we propose modifying the initial nuclear structure. As previously discussed, anisotropic flow is indicative of the initial spatial anisotropies in the overlap region of colliding nuclei. Therefore, we aim to mitigate this issue by reducing initial state fluctuations, specifically by altering the minimum distance between nucleons within the Pb nucleus. Additionally, shear viscosity has the potential to suppress the coefficient of anisotropic flow. Consequently, it is necessary to adopt a shear viscosity coefficient that is lower than that predicted by the Woods-Saxon distribution.

In this paper, we focus on decreasing the initial fluctuations and taking a smaller viscosity than the Woods-Saxon distribution to study $v_2\{2\}$ and $v_3\{2\}$ puzzle . We find that decreasing the initial fluctuations can improve $v_2\{2\}$ and $v_3\{2\}$ puzzle. The article is organized as follows: Section II introduces the initial state model and hydrodynamics model. Section III presents the results of initial state eccentricities and flow harmonic coefficients. Section IV summarizes the results.

\section{methods}

\subsection{Initial state geometry}

The atomic nucleus, composed of protons and neutrons, is considered one of the most complex quantum-mechanical systems in nature. The spatial distribution of protons and neutrons within the nucleus is typically described by a Woods-Saxon distribution, given by

\begin{equation}
    \rho(r) = \frac{\rho_0}{1 + exp(\frac{r-R}{a})}.
\end{equation}
where R = 6.62 fm and a = 0.546 fm are the parameters for a Pb nucleus, $\rho_{0}$ is the saturated nuclear density.

To simulate the initial state geometry of nuclear collisions, various models have been proposed. One such model is the Reduced Thickness Event-by-event Nuclear Topology (TRENTo) \cite{PhysRevC.92.011901}, which is a simple, fast model for the initial conditions of high-energy nuclear collisions, developed by Duke University. It can additionally describe higher-order anisotropic flow $v_{n}$ due to the inclusion of entropy or energy density fluctuations in the transverse plane. Since TRENTo is very flexible and successful, this is used as the default for the public version of CLVisc. 

TRENTo introduces a parametric initial-condition model for high-energy nuclear collisions, based on eikonal entropy deposition via a "reduced-thickness" function:
\begin{equation}
    f = T_{R}(p; T_{A}, T_{B}) = (\frac{T^{p}_{A} + T^{p}_{B}}{2} )^{1/p},
\end{equation}
where $T_{A,B}(x,y) = \int dz \rho^{part}_{A,B}(x,y,z)$. The TRENTo model has successfully described experimental proton-proton, proton-nucleus, and nucleus-nucleus multiplicity distributions, and it has generated nucleus-nucleus eccentricity harmonics consistent with experimental flow constraints.
The root mean square eccentricities $\epsilon_{n}\{2\}$ can be expressed as 

\begin{equation}
    \epsilon _{n} \{ 2 \} = \sqrt{ \langle \epsilon_{n} \epsilon_{n}^{\ast}\rangle_{ \rm events}}, \epsilon_{n} = \langle r^{n} e^{in \phi}\rangle ,
\end{equation}
which describes the initial state profile of the TRENTo model. Eccentricity harmonics $\epsilon_{n}$ are calculated by using the definition,
\begin{equation}
    \langle \epsilon_{n} \rangle = \frac{\int dx dy r^{n} e^{in \phi} T_{R}}{\int dx dy r^{n} T_{R} }
\end{equation}
where the $T_{R}$ is reduced thickness function from the TRENTo model. 

\subsection{The uniformity of nucleon distribution within nucleus}

\begin{figure}
    \centering
    \includegraphics[width=0.49\textwidth]{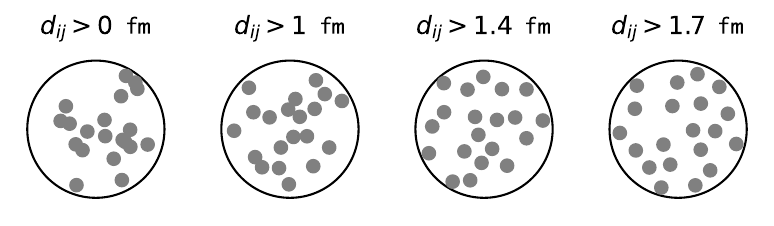}
    \caption{A schematic diagram illustrating the relationship between the uniformity of nucleon distribution within the nucleus and the minimum distance between nucleons.}
    \label{fig:uniformity}
\end{figure}

Inspired by \cite{Denicol:2014ywa}, where nucleon-nucleon correlation plays a role in resolving the $v_2$ to $v_3$ ratio puzzle in ultra-central collisions, we hypothesize that this effect arises because the short-range repulsion introduced by nucleon-nucleon correlation leads to a more uniform distribution of nucleons within the nucleus. We pose the question: Could a uniformly distributed nucleon configuration within the nucleus fully resolve this puzzle? To address this, we manipulate the uniformity of nucleon distribution by varying the minimum distances between nucleons. Given that the single nucleon density and the charge radius of the nucleus are experimentally constrained, increasing the minimum distance between nucleons enhances the uniformity of the nucleon distribution. This is illustrated in Fig.~\ref{fig:uniformity}, where 20 nucleons are sampled within a 2-dimensional circle of radius 5 fm. The same rejection method as in TRENTo is employed to discard newly sampled nucleons if their distance from previously sampled nucleons is less than the minimum distance $d_{ij}$. It is shown that with $d_{ij}=0$ fm, the nucleons exhibit strong fluctuations and significant overlaps, which are unfavorable under the short-range repulsive potential. As $d_{ij}$ increases from 1 fm to 1.7 fm, the nucleons distribute more uniformly inside the nucleus. We aim to test the impact of this uniformity on the $v_2$ and $v_3$ ratios in ultra-central collisions.

Noting that this is a proof-of-principle study, the minimum nucleon separation does not imply that nucleons cannot be closer than this distance in reality. Short-range correlations and the EMC effect have already demonstrated that there are nucleon pairs with large relative momenta and short relative distances within nuclei. As shown in Figure~\ref{fig:uniformity}, the minimum distance parameter provides a convenient method to alter the homogeneity of nucleons within the nucleus.




\subsection{Sub-nucleon structure}

The sub-nucleon structure may affect the uniformity of the nucleon distribution. Specifically, proton structure has been experimentally observed to exhibit internal "hot spots", characterized by localized regions of high gluon density \cite{DEROECK199261}. By incorporating this effect, the model offers a more realistic and detailed description of the initial state, which may enhance the predictive power of subsequent hydrodynamic simulations.  

Within the framework of sub-nucleon studies, it is commonly assumed that a nucleon possesses $N_{c}$ degrees of freedom. These degrees of freedom partition the nucleon into $N_{c}$ Gaussian components. For the specific case of three constituent quarks, $N_{c}$ is set to 3
\cite{PhysRevC.94.024907,PhysRevC.94.024914,MANTYSAARI2017832,Mantysaari:2023gop,CPC:10.1088/1674-1137/ada7d1}. The spatial coordinates of these constituent quarks within the nucleon are sampled from a Normal distribution,
\begin{equation}
    p(x,y) = \frac{1}{2 \pi r^2} \exp [-\frac{(x-x'^2)+ (y-y'^2)}{2 r^2}],
\end{equation}
where $(x', y')$ denotes the transverse position of the parent nucleon, and r represents the constituent dispersion width, calculated as follows, 
\begin{equation}
    r = \sqrt{\frac{w^{2} - v^{2}}{1-1/N_{c}}} .
\end{equation}
where $w$ is the Gaussian width of a nucleon fixed at 0.5 fm, and $v$ denotes the Gaussian width of a constituent quark, whose value is set to 0.3 fm. These are the default settings in Trento.

\begin{figure}
    \centering
    \includegraphics[width=0.45\textwidth]{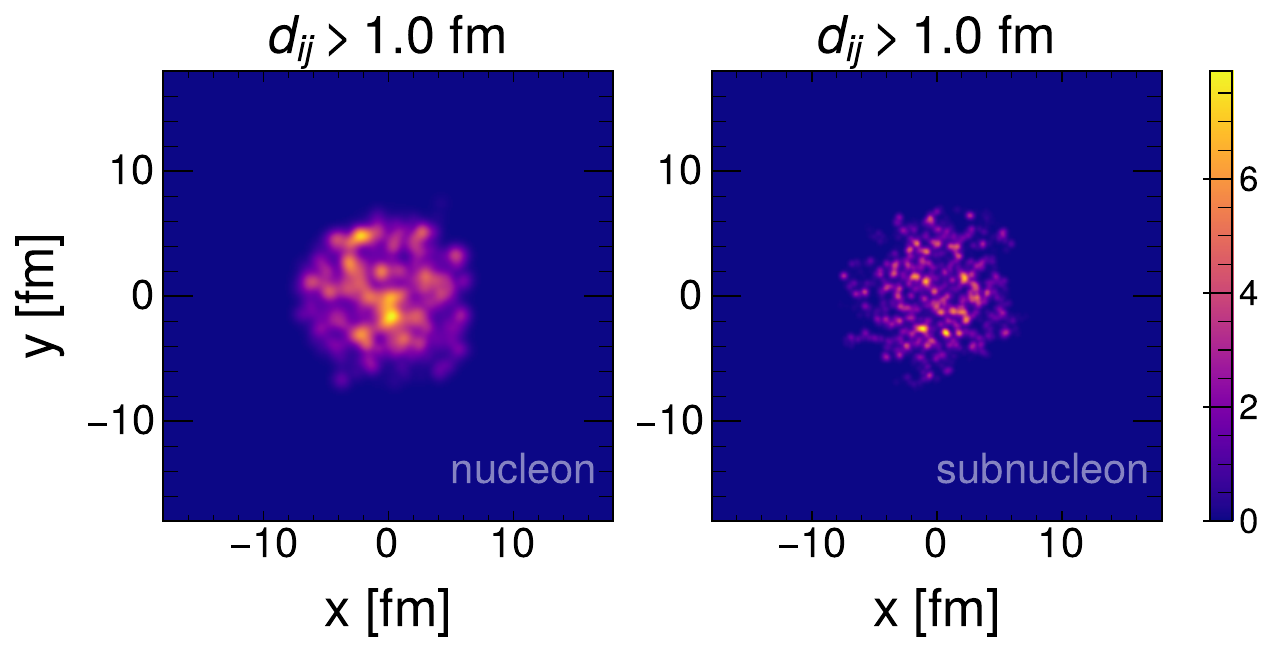}
    \caption{The distribution of initial entropy density in the transverse plane for Pb + Pb collisions at $\sqrt{s_{\mathrm{NN}}}$ = 2.76 TeV, with and without sub-nucleon fluctuations.}
    \label{fig:entropy-density}
\end{figure}

Fig.~\ref{fig:entropy-density} compares the initial entropy density distributions in the transverse plane for two ultra-central Pb+Pb collisions, with and without sub-nucleon fluctuations. The distributions exhibit significant visual differences. In the absence of sub-nucleon fluctuations, the entropy density distribution appears more diffused. Conversely, when sub-nucleon fluctuations are included, the entropy density distribution becomes notably sharper. The size of each hot spot is considerably smaller in the presence of sub-nucleon fluctuations compared to the case without them. These observations suggest that the sub-nucleon structure influences the initial state fluctuations. However, it remains uncertain whether it also affects the uniformity of the distribution.

\subsection{Hydrodynamics simulation}
We investigate the hydrodynamic response to these initial conditions using CLVisc \cite{PhysRevC.97.064918,PANG2016272,Pang:2015zrq,PhysRevC.91.044904,PhysRevC.82.014903,PhysRevC.95.014906,Wu:2021fjf} a (3+1)-dimensional relativistic hydrodynamic model parallelized on graphics processing units (GPUs) using the Open Computing Language (OpenCL). This implementation achieves a 60-fold performance enhancement in spacetime evolution and over a 120-fold improvement in Cooper-Frye particlization compared to non-GPU parallelized computations. The model's validity is rigorously assessed through comparisons with various analytical solutions, existing numerical hydrodynamic approaches, and experimental data on hadron spectra from high-energy heavy-ion collisions.

The main equations of hydrodynamics are derived from the conservation laws of energy and momentum, expressed as:
\begin{equation}
    \nabla_{\mu} T^{\mu \nu} = 0 ,
\end{equation}
 where the energy-momentum tensor ($T^{\mu \nu}$) is defined as: 
 $T^{\mu \nu} = \epsilon u^{\mu}u^{\nu} - P \Delta ^{\mu \nu} + \pi^{\mu \nu}$. Here  $\epsilon$, $P$, $u^{\mu}$ and $\pi^{\mu \nu}$ represent the energy density, pressure, four-velocity, and shear-stress tensor respectively. The projection operator($\Delta^{\mu \nu}$) is given by: $\Delta^{\mu \nu} = g^{\mu \nu} -u^{\mu}u^{\nu}$. 

The dynamical models based on relativistic hydrodynamics have been successful in describing the anisotropic flow coefficients $ v_{n}$. These quantities also are conveniently represented by the 'flow vector' $V_{n} = v_{n} \exp({i n \Phi_{n}})$ in each part. The $V_{n}$ value reflects the hydrodynamic response of the produced medium to the $n^{th}$ -order initial-state eccentricity vector \cite{PhysRevC.85.024908}, denoted by $\epsilon_{n} = \epsilon_{n} \exp({i n \psi_{n}})$. Model calculations show that $V_{n}$ is approximately proportional to $\epsilon_{n}$ in general for n = 2 and 3, and for n = 4 in the case of central collisions.

We compute the root mean square of the flow harmonic coefficients $V_{n}$,
\begin{equation}
    v_{n}\{2\} = \sqrt{\langle V_{n} V_{n}^{*}\rangle_{events}},
    V_{n} = \langle exp(i n \phi) \rangle,
\end{equation}
where $\phi$ is the azimuthal transverse momentum angle, $\langle ... \rangle_{events}$ is the average for events, and $V_n$ is expressed as 
\begin{equation}
    \langle ... \rangle = \frac{\int d^{2} p_{T}  \frac{dN}{2 \pi p_{T}dp_{T}} (...)}{\int d^{2} p_{T}  \frac{dN}{2 \pi p_{T}dp_{T} }},
\end{equation}
where the $\frac{dN}{2 \pi p_{T}dp_{T} d \eta }$ is the particle distributions in the transverse momentum range at the central rapidity zone.

In our computational framework, the lattice dimensions are configured as \( n_{\rm x} \times n_{ \rm y} \times n_{\rm z} = 300 \times 300 \times 161 \), with a thermalization time of \( \tau_{0} = 0.6 \) fm. The equation of state utilized in our simulations is employed in "lattice-{pce165}". Additionally, we adopted a freeze-out temperature of T = 0.137 GeV. For each configuration of the model parameters, we generated 1000 hydrodynamic events. 

\section{Results}

In this section, we present the results of our study, which involves calibrating parameters using pseudo-rapidity distributions, transverse momentum, and various flow coefficients for charged hadrons in Pb + Pb collisions at $\sqrt{s_{\mathrm{NN}}}$ = 2.76 TeV. We then examine the initial eccentricity for different minimal distances at the 0-1$\%$ centrality using the TRENTo model. Finally, we present numerical results for flow harmonics across varying minimal distances and centrality ranges using the TRENTo initial condition and CLVisc hydrodynamics model.


\subsection{Code calibration}


\begin{figure}
    \centering
    \includegraphics[width=0.45\textwidth]{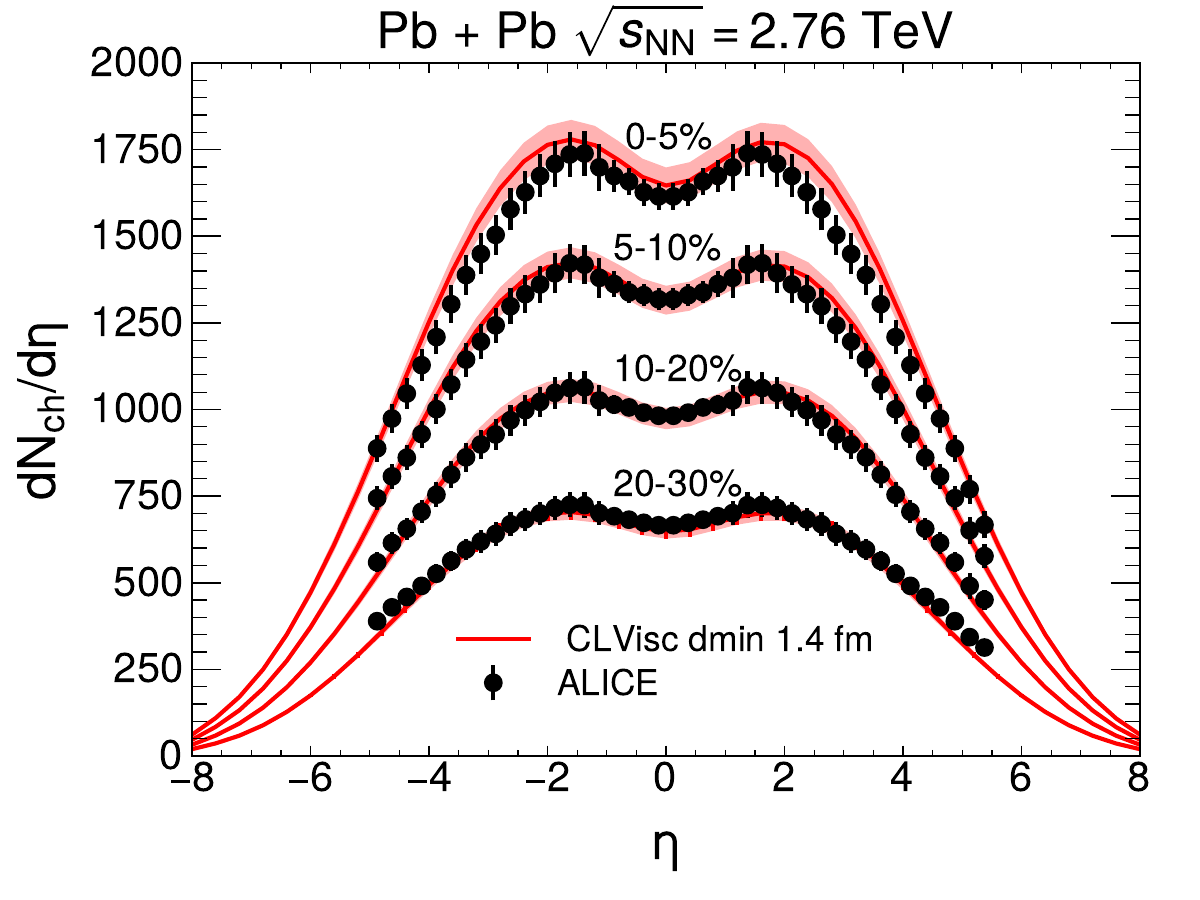}
    \caption{Pseudorapidity distribution for charged hadron in Pb + Pb collisions at $\sqrt{s_{\mathrm{NN}}}$ =2.76 TeV with centrality range 0-5 $\%$, 5-10 $\%$, 10-20 $\%$ and 20-30 $\%$, from CLVisc (the shear viscosity $\eta /s$ = 0.16) and LHC experimental data by the ALICE collaboration \cite{2017567}.}
    \label{fig:enter-label1}
\end{figure}

Fig.\ref{fig:enter-label1} (from upper to lower) presents the pseudo-rapidity distributions of charged hadrons in four distinct centrality classes, 0-5$\%$, 5-10$\%$, 10-20$\%$ and 20-30$\%$, obtained from CLVisc simulations, compared with experimental data from the ALICE collaboration. The simulation results, with a minimum distance between nucleon pairs $ d_{\rm min}=1.4$ fm, exhibit good agreement with the experimental data. This comparison serves to calibrate model parameters in relativistic hydrodynamic simulations for the most central collisions. Initially, we hypothesized that varying $d_{\rm min}$ might alter the centrality dependence, thereby undermining the predictive power of relativistic hydrodynamics for other centrality ranges when parameters are fixed using the 0-5\% centrality range, as observed in O-O collisions considering $\alpha$ clusters \cite{Ding_2023}. Contrary to our hypothesis, $d_{\rm min}$ does not influence the centrality dependence. This outcome may be attributed to the fact that \(\alpha\) clusters induce more substantial changes to the nuclear structure of \(\rm ^{16}O\) than the modifications we applied to \(\rm ^{208}Pb\).

\begin{figure}
    \centering
    \includegraphics[width=0.45\textwidth]{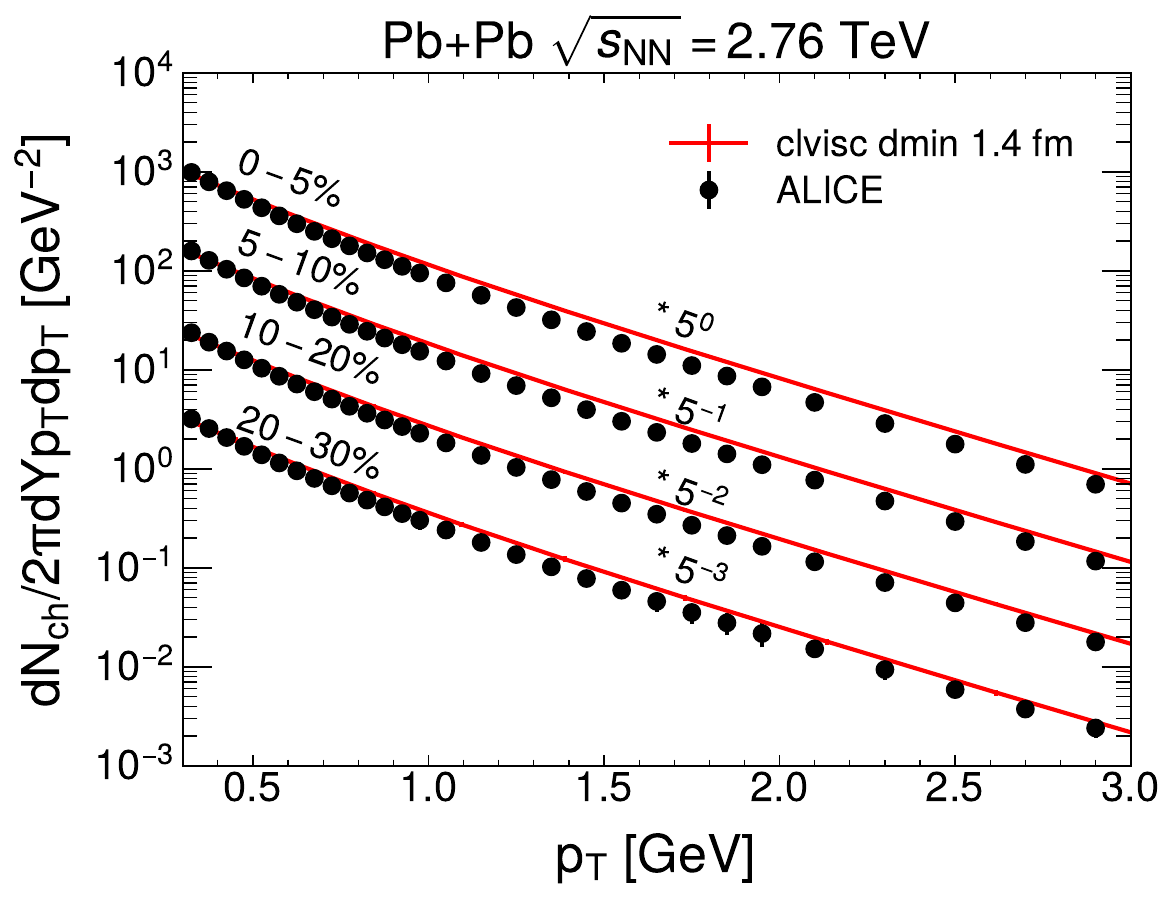}
    \caption{The transverse momentum spectra of charged hadron in Pb + Pb collision at $\sqrt{s_{\mathrm{NN}}}$ = 2.76 TeV. Comparison of CLVisc hydrodynamic model calculations (curves) with the shear viscosity $\eta /s$ = 0.16 with ALICE experimental measurements (points) \cite{ALICE:2012aqc} for four centrality classes: 0-5$\%$, 5-10$\%$, 10-20$\%$ and 20-30$\%$.}
    \label{fig:enter-label6}
\end{figure}

Fig.\ref{fig:enter-label6} shows the transverse momentum spectra of charged hadrons as compared with experimental data, for four distinct centrality classes, 0-5 $\%$, 5-10 $\%$, 10-20 $\%$ and 20-30 $\%$. The red lines correspond to CLVisc simulations with $\eta / s = 0.16$ and the black dots represent experimental data from the ALICE collaboration. Our observations indicate that a larger shear viscosity coefficient ($\eta/s $) suppresses the high-momentum hadron spectrum, making it more consistent with the experimental data. However, hydrodynamic simulations tend to overestimate the high transverse momentum charged hadron spectra compared to the experimental data from the ALICE Collaboration, in order to describe the anisotropic flows. This suggests that incorporating bulk viscosity into the 3+1-dimensional viscous hydrodynamics framework may be necessary to improve the accuracy of the simulations.
\begin{figure*}
    \centering
    \includegraphics[width=0.85\textwidth]{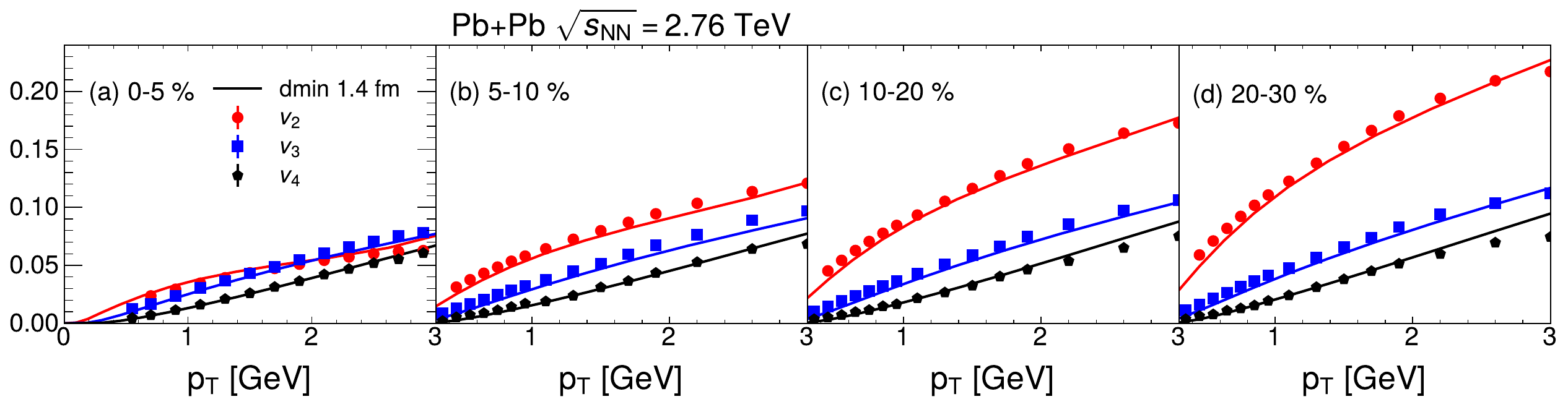}
    \caption{Comparison of anisotropic flow coefficients $v_2$, $v_3$ and $v_4$ obtained via the event plane method in Pb+Pb collisions at $\sqrt{s_{\mathrm{NN}}}$ =2.76 TeV for centrality classes 0-5$\%$,5-10$\%$, 10-20$\%$, 20-30$\%$. Results from CLVisc hydrodynamic simulations (solid curves) with shear viscosity $\eta /s$ = 0.16 are compared with experimental data (points) from the ATLAS Collaboration \cite{ATLAS:2012at}.}
    \label{fig:enter-label3}
\end{figure*}

Fig.\ref{fig:enter-label3} presents a comparative analysis of high-order differential flow between CLVisc simulations and ALICE experimental data across four centrality ranges 0-5$\%$, 5-10$\%$, 10-20$\%$, 20-30$\%$. The figure sequentially displays differential flow coefficients $v_2$, $v_3$, $v_4$ from top to down at small $p_T$. These results were derived using a Woods-Saxon distribution with a minimum distance parameter of 1.4 fm. The simulation results demonstrate that the minimum distances for flow coefficients $v_2$ and $v_4$ show good agreement with experimental data. Our analysis indicates that relativistic hydrodynamic simulations effectively describe the $p_T$ differential $v_n$ for semi-central collisions. However, the ultra-central puzzle requires consideration of $p_T$ integrated $v_n$. For semi-central collisions, our $p_T$ spectra exhibit an excess at high $p_T$ values, which may introduce minor discrepancies in the $p_T$ integrated $v_n$, particularly for the 0-5\% collision centrality. Notably, the $p_T$ differential $v_3$ is larger than $v_2$ at $p_T>2$ GeV for 0-5\% centrality. 



\subsection{Initial eccentricities}

\begin{figure}
    \centering
    \includegraphics[width=0.45\textwidth]{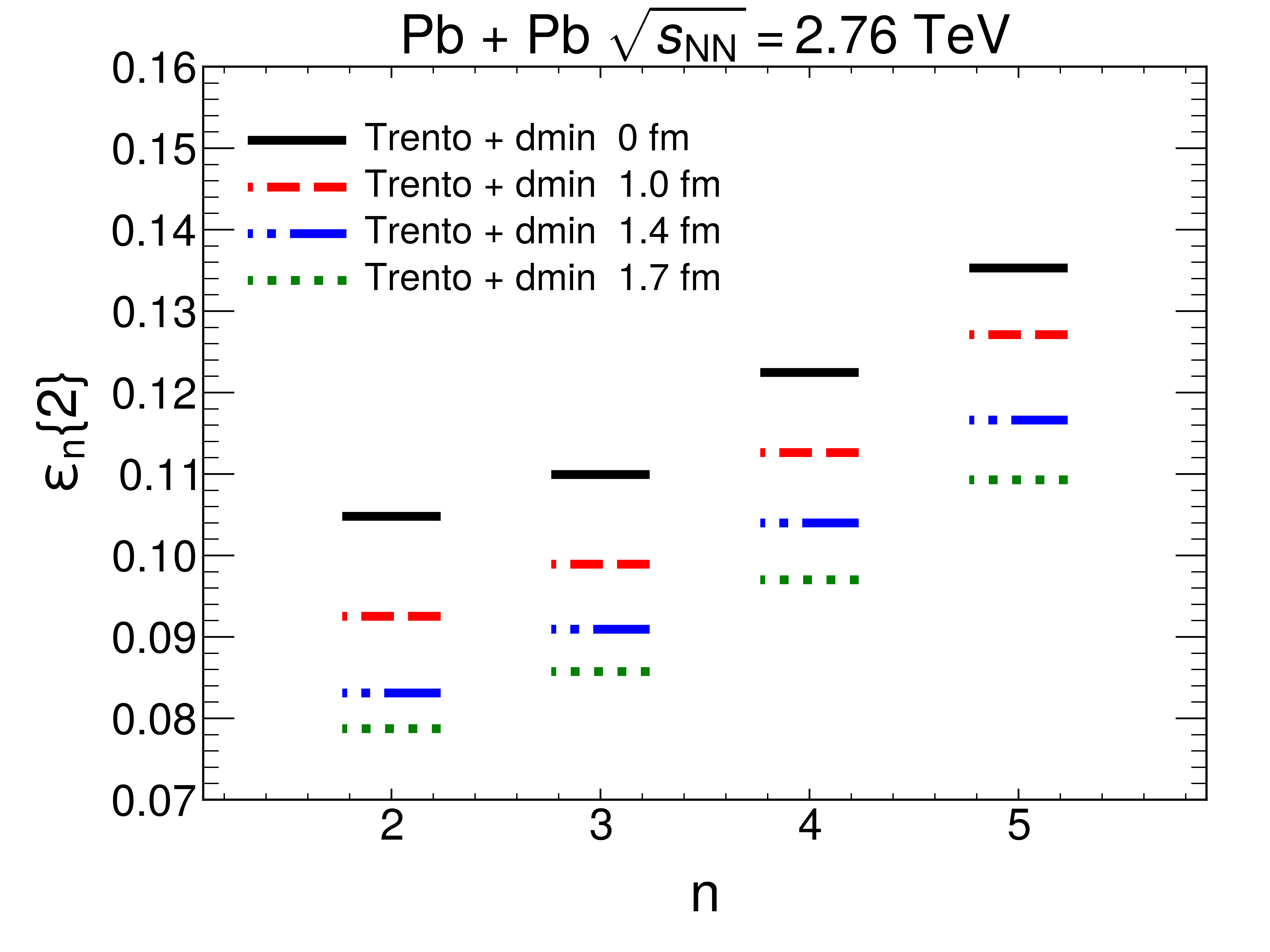}
    \caption{(Color online) The root mean square of eccentricities $\epsilon_{n}\{2\}$ calculated using the TRENTo initial condition model for 0-1\% centrality Pb+Pb collisions. The black line is minimum distance of nucleon-nucleon dmin = 0.0 fm (solid line), dmin = 1.0 fm (red dashed line), dmin = 1.4 fm (blue dash-dot line) and dmin = 1.7 fm (green dotted line).}
    \label{fig:enter-label}
\end{figure}

\begin{figure}
    \centering
    \includegraphics[width=0.45\textwidth]{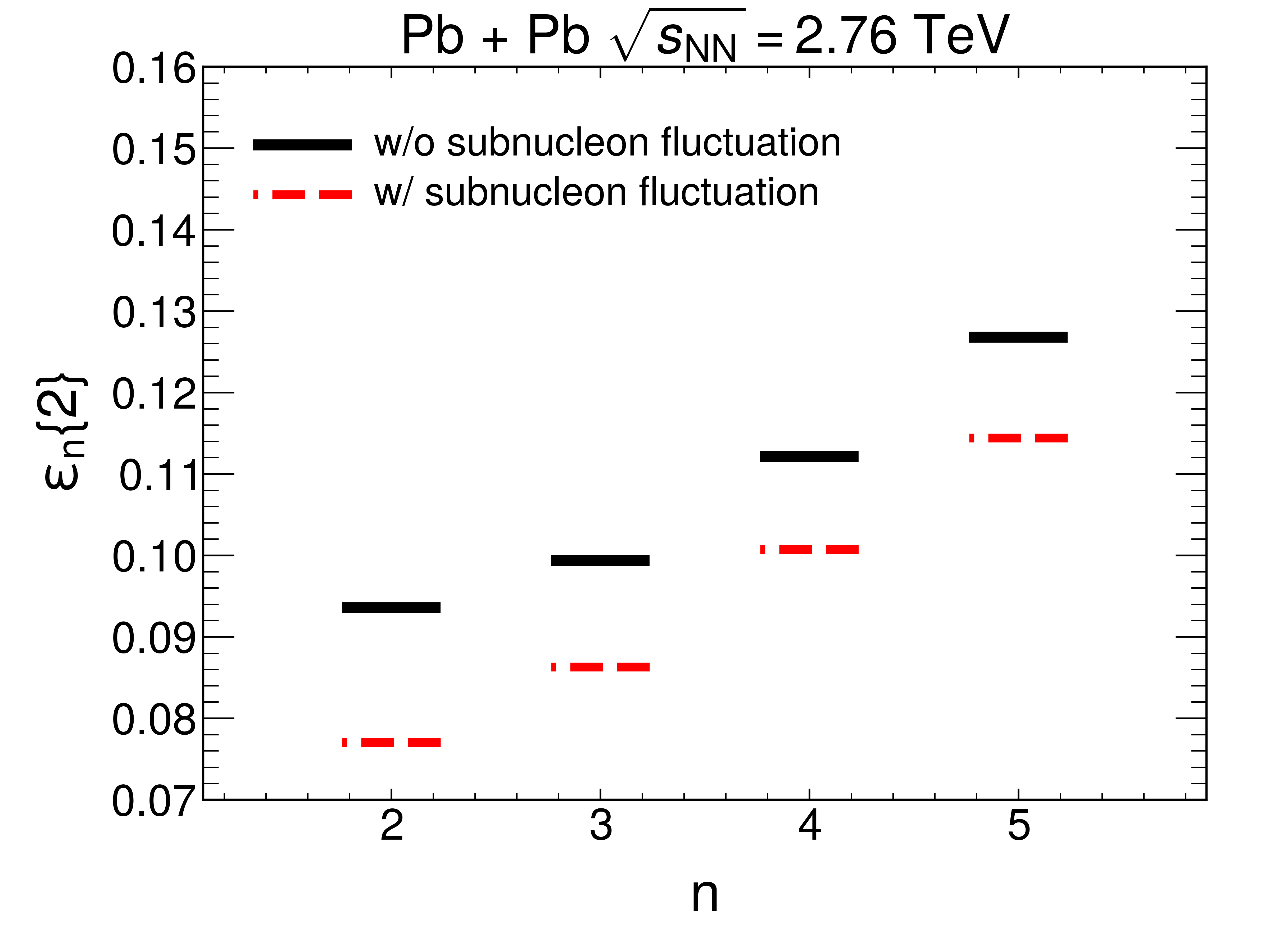}
    \caption{The root mean square of eccentricities $\epsilon_{n}\{2\}$ in 0-1\% central Pb+Pb collisions calculated using the TRENTo initial condition model. The black solid line(dmin = 1.0 fm) and red dashed line(dmin = 1.0 fm and sub-nucleon fluctuations) represent the model predictions.}
    \label{fig:enter-label9}
\end{figure}

The initial eccentricity exhibits an approximate linear relationship with the final collective flow that $v_n = k \epsilon_n$. Therefore, we first analyze the influence of the minimum distance on the initial eccentricity. The centrality of the collisions was determined using the "mult" parameter from the TRENTo model. For example, to select events within the 0–5$\%$ centrality range, all events were sorted based on the "mult" parameter in descending order, and the top 5$\%$ of events were selected. Fig.\ref{fig:enter-label} illustrates the root mean square eccentricities $\epsilon_{n} \{2 \}$, for n=2-6, in Pb-Pb collisions within the 0-1$\%$ centrality class, as a function of the minimum distance from the nucleus. The data reveal that as the minimum distance increases, the root mean square eccentricities exhibit a gradual decrease, with the rate of decrease remaining approximately constant. Notably, higher-order flows are significantly larger than the second-order flows. The results are presented for the TRENTo model, which indicates that an increase in the minimum distance between nucleons correlates with a reduction in fluctuations in the initial conditions, where the eccentricity primarily arises from fluctuations in the initial nucleon states. Secondly, we consider the effect of sub-nucleon fluctuations, as illustrated in Fig.\ref{fig:enter-label9}. We found that sub-nucleon fluctuations can also reduce the initial eccentricity. 

\subsection{Flow harmonic coefficients}

\begin{figure}
    \centering
    \includegraphics[width=0.45\textwidth]{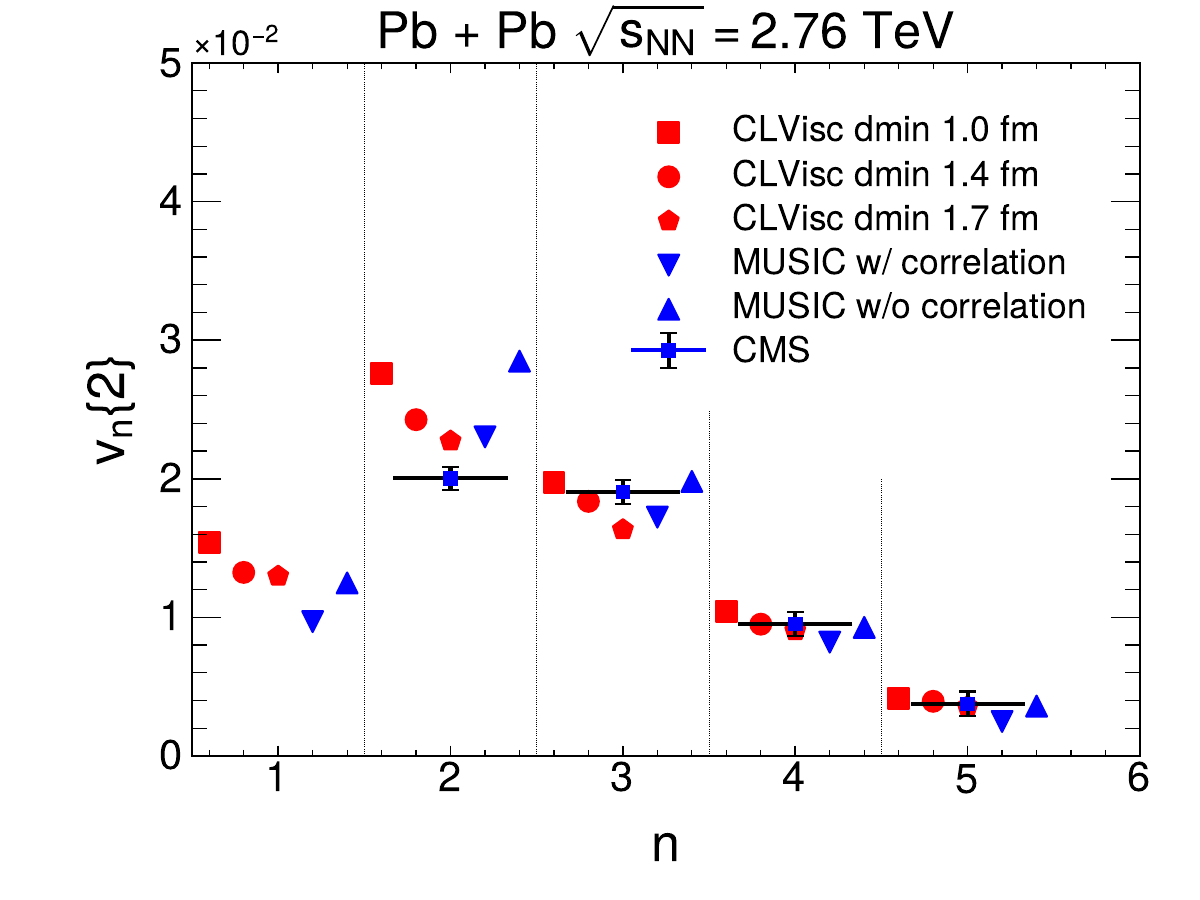}
    \caption{The flow harmonics, $v_{n} \{2\}$, for the TRENTo initial conditions and CLVisc hydrodynamics models are presented. The red square corresponds to the minimum distance ( dmin = 1.0 fm ), the red circle ( dmin = 1.4 fm ), and the red pentagon ( dmin = 1.7 fm ). Additionally, the blue upper triangle represents the scenario without nucleon-nucleon correlations, while the blue down triangle corresponds to the case with nucleon-nucleon correlations, both derived from the IP-glasma + MUSIC model \cite{Denicol:2014ywa}.}
    \label{fig:enter-label4}
\end{figure}


Fig.\ref{fig:enter-label4} presents the flow harmonics coefficients $v_{n} \{2 \}$ for minimum distances d = 1.0, 1.4, and 1.7 fm, considering both the nuclear-nuclear correlation (MUSIC + IP-glasma) and without nuclear-nuclear correlation, for (n = 2-5 ), in Pb-Pb collisions at $\sqrt{s_{\mathrm{NN}}}$ = 2.76 TeV within the 0-1$\%$ centrality class, for charged hadrons. Notably, compared to the 0-5$\%$ centrality class, the more central 0-1$\%$ centrality class in Pb-Pb collisions appears to require a higher shear viscosity to entropy density ratio $ \eta / s$. Consequently, a value of $ \eta/s $ = 0.22 was employed in this analysis. To facilitate comparison with experimental data, a cut-off of 0.3-3 GeV was applied to the transverse momentum integrals. We find that the result of the minimum distance $d_{\rm min}$ = 1.4 fm improves the previous results, leading to a better agreement with the CMS data \cite{CMS:2013bza}. 
\begin{figure*}
    \centering
    \includegraphics[width=0.95\linewidth, height=5cm]{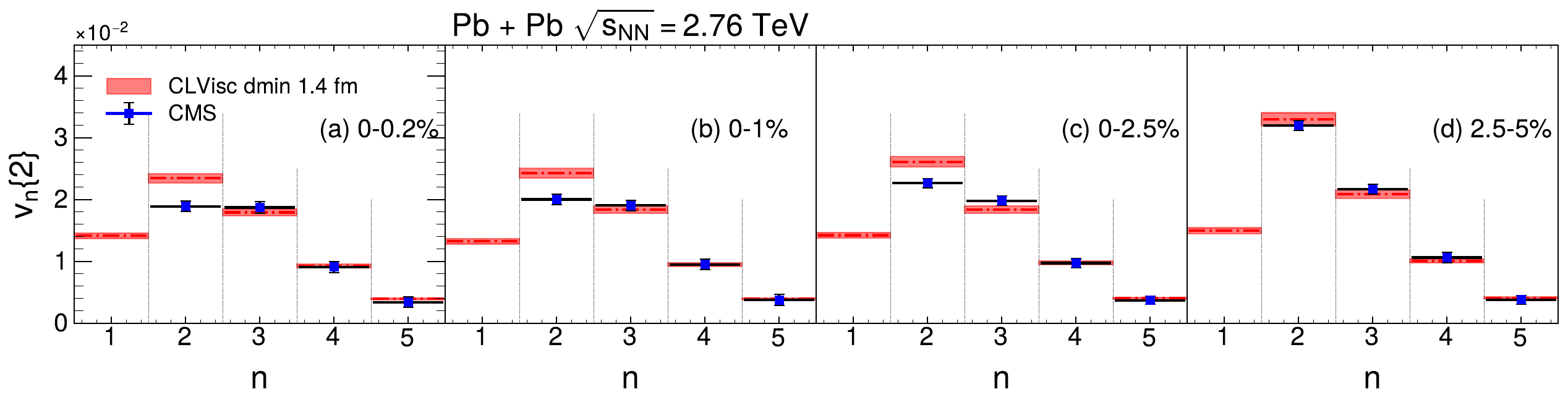}
    \caption{(Color online) The flow harmonics, $v_{n} \{2\}$, calculated using TRENTo initial conditions coupled to CLVisc hydrodynamic simulations with shear viscosity $\eta /s$ =0.22, compared with CMS experimental data. Results are shown for four centrality classes:0-0.2$\%$, 0-1$\%$, 0-2.5$\%$, 2.5-5$\%$. The red line displays model predictions using a minimum nucleon-nucleon distance dmin = 1.4 fm, while black line represent CMS experimental measurements. Vertical error bars indicate statistical uncertainties in experimental data, with shaded color bands corresponding to theoretical statistical uncertainties from the model calculations.}
    \label{fig:enter-label5}
\end{figure*}

Fig.\ref{fig:enter-label5} shows comparison between hydrodynamic simulation results with centrality of 0-0.2$\%$, 0-1$\%$, 0-2.5$\%$, 2.5-5$\%$ and experimental data from CMS \cite{CMS:2013bza}. And we find that the results from fluid dynamics are nearly consistent with experimental results for centrality ranges of 2.5-5$\%$. However, for other centralities, the minimization of the distance improved the consistency between the hydrodynamics results and experimental values of $v_{2} \{2\}$ and $v_{3} \{2\}$. For other higher-order flows, the hydrodynamic simulation aligns well with the experimental observations.
\begin{figure}
    \centering
    \includegraphics[width=0.45\textwidth]{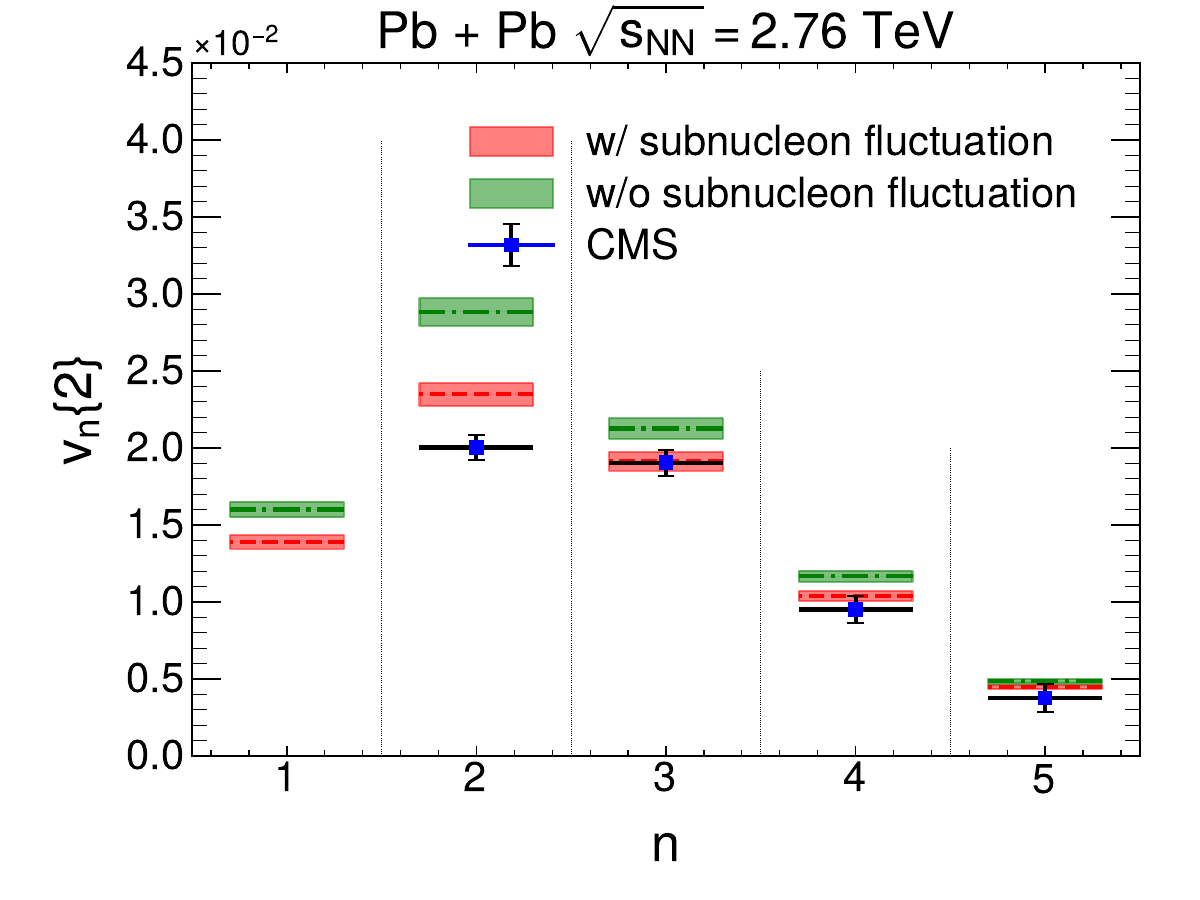}
    \caption{Comparison of the flow coefficient $v_{n} \{2 \}$ between experimental data and theoretical predictions from the dmin = 1.0 fm and sub-nucleon fluctuation with dmin =1.0 fm for Pb+Pb collisions at  $\sqrt{s_{\mathrm{NN}}}$ = 2.76 TeV within the 0-1$\%$ centrality range, from CLVisc with $\eta /s$ =0.18.  The error bar represents the statistical error on the experiment, and the shaded color boxes represent the theoretical statistical error.}
    \label{fig:enter-label8}
\end{figure}

Fig.\ref{fig:enter-label8} illustrates the influence of sub-nucleon fluctuations on flow coefficients in ultra-central collisions. A comparison of experimental data with calculations from two different initial condition models, including scenarios with and without sub-nucleon fluctuations $d_{\text{min}} = 1.0$ fm, provides a clear evaluation of the impact of sub-nucleon fluctuations on the hydrodynamic final state. For Pb+Pb collisions at $\sqrt{s_{\mathrm{NN}}} = 2.76$ TeV within the 0-1\% centrality range, the predictions for the second-order flow coefficient $v_{2} \{2 \}$ and the ratio of $v_{2} \{2 \}$ to $v_{2} \{2 \}$ demonstrate that results incorporating sub-nucleon fluctuations align more closely with experimental observations than those without. The exclusion of sub-nucleon fluctuations in the $d_{\text{min}} = 1.4$ fm case is justified by the fact that $v_{3} \{2 \}$ and $v_{4} \{2 \}$ values would significantly underestimate the experimental data. For the same reason, a smaller $\eta/s$ has been used as compared with Fig.\ref{fig:enter-label5}.



\section{Discussion and summary}


The $v_{2} \{2\}$ to $v_{3} \{2\}$ ratio puzzle in ultra-central heavy ion collisions remains an unresolved challenge. Previous studies have shown that modifying the two-nucleon distribution within the colliding nucleus can bring the $v_{2} \{2\}$ to $v_{3} \{2\}$ ratio closer to experimental data, but this modification has not fully resolved the puzzle. We hypothesize that this discrepancy may be due to more intricate nuclear structure effects within the nucleus, which render the nucleon distribution within the nucleus inadequately represented by a simple Woods-Saxon distribution. To explore this possibility, we conducted a preliminary investigation by altering the minimum distance between nucleons within the nucleus and incorporating sub-nucleon fluctuations within nucleon to modify the nuclear structure and observe its impact on the $v_{2} \{2\}$ to $v_{3} \{2\}$ ratio.

We employed the TRENTo model to sample the nucleon distribution within the nucleus and utilized the 3 + 1 dimensional hydrodynamics code CLVisc to simulate Pb+Pb collisions at a center-of-mass energy of $\sqrt{s_{\mathrm{NN}}}=2.76$ TeV, capturing the spatiotemporal evolution of the quark-gluon plasma. Our findings indicate that increasing the minimum distance between nucleons reduces the geometric eccentricity of all orders. Consequently, this reduction allows for a smaller shear viscosity to entropy density ratio $(\eta/s)$ in relativistic hydrodynamic simulations to achieve a $v_{2} \{2\}$ to $v_{3} \{2\}$ ratio that aligns more closely with experimental data. Without increasing the minimum nucleon distance, the higher geometric eccentricities necessitate a larger $\eta/S$ to match the experimental $v_{2} \{2\}$, which in turn leads to a stronger suppression of higher-order anisotropic flows.

Our study suggests that nuclear structure indeed plays a role in resolving the $v_{2} \{2\}$ to $v_{3} \{2\}$ ratio puzzle in ultra-central heavy ion collisions. However, this resolution is intertwined with other effects, such as the reduction in the required shear viscosity, which mitigates the damping effect of viscosity on higher-order anisotropies. It is important to note that our current research focuses on one aspect of the puzzle and does not comprehensively consider other factors that may influence it, such as bulk viscosity and hydrodynamic fluctuations. A thorough resolution of the puzzle will require future studies that integrate all relevant effects. 

Additionally, our findings indicate that sub-nucleon fluctuations also play a significant role in addressing the $v_{2} \{2\}$ to $v_{3} \{2\}$ ratio puzzle in ultra-central heavy ion collisions. This phenomenon can be attributed to the enhanced spatial homogeneity of the quark-gluon plasma initial geometry. Considering sub-nucleon effects, the initial eccentricity decreases, with the second-order eccentricity decreasing more than the third-order eccentricity, which offers substantial explanatory power for resolving this longstanding puzzle in collective flow phenomenology.


The influence of pileup events to $v_{2} \{2\}$ to $v_{3} \{2\}$ ratio puzzle in ultra-central events is proved to be negligible  \cite{CMS:2013bza}. Pileup events, which occur when two collisions are recorded within a single beam crossing, are typically identified by experimentalists through the correlation of energy sum signals between the Zero Degree Calorimeter (ZDC) and Hadron Forward (HF) detectors. Events with large signals are labeled as pileup events, which constitute approximately 0.1\% of all events. These events are subsequently rejected, and the results remain stable within less than 1\% when varying the ZDC sum energy requirements for pileup rejection, as demonstrated in the CMS experiment \cite{CMS:2013bza}.

Furthermore, our findings underscore the sensitivity of ultra-central heavy-ion collision events to nuclear structure. The unresolved puzzles in heavy-ion collisions may necessitate interdisciplinary research that bridges high-energy nuclear physics with low-energy nuclear physics to jointly address these challenges and drive progress in both fields.

In conclusion, while our study provides insights into the $v_{2} \{2\}$ to $v_{3} \{2\}$ ratio puzzle, a comprehensive understanding remains an ongoing endeavor that will benefit from a synergistic approach across different areas of nuclear physics.

\section{Acknowledgments}
This work is supported by the National Natural Science Foundation of China under Grant No.\ 12075098, No.\ 12435009 and No.\ 1193507, and the Guang-dong MPBAR with No.2020B0301030008. The numerical calculation have been performed on the GPU cluster in the Nuclear Science Computing Center at Central China Normal University (NSC3).

\end{document}